# Connecting disc-corona physics and ionised outflows in AGN in the 2040s

Susanna Bisogni[1], Giustina Vietri[1], Andrea Travascio[2], Adriana Gargiulo[1], Chiara Mancini[1] and Swayamtrupta Panda[3]

[1] INAF – Istituto di Astrofisica Spaziale e Fisica cosmica Milano, Via Alfonso Corti 12, 20133 Milano, Italy
[2] INAF – Osservatorio Astronomico di Trieste, Via G. B. Tiepolo 11, I-34131 Trieste, Italy
[3] International Gemini Observatory/NSF NOIRLab, La Serena, Chile

**From disc-corona coupling to galaxy-scale feedback.** Tight correlations between supermassive black hole (SMBH) mass and host-galaxy properties suggest a close co-evolution (Kormendy&Ho 2013), but the physical processes that connect sub-pc accretion flows to kpc-scale gas remain poorly understood. Theoretical models of galaxy evolution invoke AGN feedback, where energy released by accretion is transported outwards by winds, outflows and radiation over many orders of magnitude in scale (Fabian 2012; Laha+21), that are indeed observed from the X-ray to sub-mm. Yet we still lack a unified empirical picture of how this energy is produced, transported, and ultimately coupled to the interstellar and circum-galactic medium (ISM/CGM).

**Central engine.** In the standard picture, viscous dissipation in an optically thick accretion disc produces thermal UV emission (Lynden-Bell 1969), while a compact hot corona Compton up-scatters disc photons into X-rays (Haardt & Maraschi 1991). Observationally, luminous unobscured ("blue") quasars follow a tight, non linear correlation between rest frame UV and X-ray luminosity, the Lx-Luv relation (e.g. Tananbaum+1979; Lusso+10, Bisogni+21, see Fig. 1). More UV-luminous AGN are, on average, relatively X-ray weaker: a tenfold increase in UV luminosity corresponds to only a factor ~4 in X-rays. The intrinsic dispersion of this relation, for carefully selected samples, is remarkably small (≤0.2 dex; Lusso&Risaliti17; Sacchi+22), and the relation does not seem to evolve with redshift (Salvestrini+19; Bisogni+21). This behaviour must reflect a universal coupling between disc and corona, yet a comprehensive physical explanation is still missing. At the same time, a significant fraction of quasars appear to be genuine outliers from the canonical Lx-Luv relation. For instance, in small, spectroscopically well-characterised samples of high-luminosity quasars, the fraction of objects intrinsically underluminous in the X-ray, compared to what is expected from the relation (non-absorbed, intrinsic X-ray weak), can reach 25-40% of the parent population, suggesting that the partition of accretion power between disc and corona is not unique and can switch between distinct accretion modes, or "coronal states" (e.g. Nardini+19; Zappacosta+20; Laurenti+22; Degli Agosti+25). More representative, statistically significant samples currently lack homogeneous X-ray spectroscopy, and are mostly limited to photometry or hardness ratios (e.g. Pu+2020, Timlin+2020). The Lx-Luv relation is often recast in terms of the optical-to-X-ray spectral index (αox), which measures how X-ray luminous a source is at a given UV luminosity. Δαox, the deviation from the expected αox, quantifies the "distance" from the relation and is used to identify X-ray weak quasars as those with Δαox below a chosen threshold.

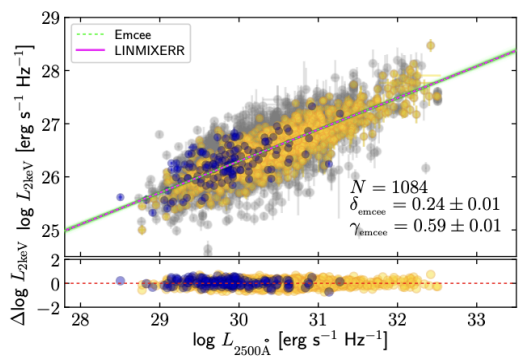

Fig. 1: The non-linear (slope~0.6) Lx-Luv relation from Bisogni+21: the dispersion for this selected sample of ~1000 sources is delta~0.24. Data are from Chandra Source Catalog 2.0/SDSS (yellow) and the COSMOS Legacy survey (blue).

**Nebular regions as calorimeters of the ionising continuum.** The EUV/very-soft-X part of the SED (~30-300 eV), the most corona-disc sensitive, is unobservable because of interstellar absorption. The gas in the broad and narrow line regions (BLR, NLR), however, is directly exposed to this radiation and reprocesses it into emission lines which, despite degeneracies with gas density, metallicity, covering factor, geometry and orientation (Ferland+20; Bisogni+17, Panda+19), act as calorimeters of the ionising continuum and encode information on the disc-corona energetics. High-ionisation lines such as C IV, He II and [O III] are especially sensitive to the soft-X/EUV continuum (e.g. Heckman+05; Temple+23); their equivalent widths and profiles also correlate with black-hole mass and Eddington ratio and are often strongly affected by winds and outflows (Marziani+17; Vietri+18; Coatman+19; Liu+24), further entangling SED and structural effects. On larger scales, extended Lyα and metal-line nebulae in the CGM, including high-ionisation emission such as C IV, provide a complementary view of how the primary central-engine radiation deposits energy and momentum into the surrounding medium (e.g. Cantalupo+14; Travascio+20) and should respond to the same high-energy SED as their BLR counterparts.

In this white paper we ask: **Are there distinct accretion states or coronal modes in AGN, and how are they imprinted on the nebular regions and the CGM?** By jointly mapping the high-ionisation lines from BLR, NLR and CGM, we aim to exploit their different sensitivities to physical conditions and geometry to break, or at least reduce, the main degeneracies and isolate the underlying ionising SED and disc-corona energetics. We propose to frame these questions in terms of the **full Lx-Luv plane**, parameterised by the offset from the canonical relation, $\Delta\alpha_{ox}$, and to use rest frame UV/optical spectroscopy as a "calorimeter" of the ionising continuum for a statistically significant sample spanning the full AGN/quasar population over a wide range of redshift, black hole mass and Eddington ratio. In parallel, JWST is revealing new AGN populations at high redshift (z>4), including X-ray-weak AGN and "little red dots" with faint high-ionisation lines, that challenge our current view of accretion physics and BLR structure at early cosmic epochs (e.g. [Yue+24](#); [Zucchi+25](#)); these sources also need to be interpreted within the broader context of disc-corona coupling.

**We propose to address the following key science questions for the 2040s**:
♦ **Are there universal accretion modes across AGN?** Does the tight Lx-Luv relation conceal a small number of distinct "states" (radiatively efficient vs inefficient coronae; compact vs extended X-ray emitting regions), or a smooth continuum of disc-corona couplings? When and why do AGN depart from the canonical relation? Are intrinsically X-ray-weak quasars, high-Eddington sources, low-Eddington/radiatively inefficient objects and changing-state AGN (e.g. [Ricci & Trakhtenbrot 2023](#)) part of a unified sequence in the Lx-Luv plane?
♦ **What are the rest frame UV/optical signatures of each accretion state?** Can we build empirical and physical scaling relations between coronal state (e.g. $\Delta\alpha_{ox}$, X-ray spectral slope) and BLR/NLR emission-line properties (strengths, ratios, profiles, asymmetries), and link the central-engine mode to the outflow properties of the nebular regions that transport energy to host-galaxy scale? To what extent can we infer the coronal state from UV/optical spectra alone, with minimal reliance on X-ray data, for large populations and at high redshift?
♦ **How do nebular regions and the CGM respond to changes in disc-corona state?** Do C IV and [O III], which both trace high-ionisation gas in BLR and NLR, respond coherently to changes in the SED, and do their profiles encode common drivers (e.g. radiation pressure, disc winds)? How does extended CGM emission (Lyα, CIV, HeII, [O III], Hα and other lines) depend on accretion state and on the time-averaged radiative output of the central engine?
♦ **How do these relations evolve with redshift and environment?** Are the newly discovered X-ray-weak, line-poor JWST AGN at high z simply the high-Eddington tail of the same underlying distribution seen at lower redshift, or do they represent qualitatively different accretion states? Can we reconcile their properties with the apparent non-evolution of the Lx-Luv relation over cosmic time?

Addressing these questions requires **large, homogeneous samples with high-quality restframe UV/optical spectroscopy and at least modest X-ray coverage**, coupled with **time-domain information** for a well-defined subsample.

**Observational strategy: the Lx-Luv plane and ensemble tomography**
We propose to use the full Lx-Luv plane, parameterised by $\Delta\alpha_{ox}$, as a **state space for accretion physics**, and to populate this plane with a **statistical sample of $10^4$-$10^5$ AGN and quasars**, the minimum size required to divide the sample into meaningful bins (50-100 object/bin) of redshift, Mbh, $\lambda_{Edd}$, to detect modest (~10-20%) systematic differences in line properties despite the large intrinsic scatter, and to obtain meaningful statistics on rare sub-populations. Such a sample can be readily extracted from the Rubin Observatory LSST, which is expected to detect ~$10^7$ AGN down to r ~ 24 ([Ivezić+19](#)). We require:
♦ **High-S/N** (S/N>3-5 per resolution element, minimum for the less luminous sources) **rest frame UV/optical spectra at R ≈ 3000-5000**, sufficient to resolve BLR/NLR profiles, measure outflow kinematics and perform accurate modelling of broad and narrow lines, including the faintest ones, such as HeII. The spectra should simultaneously cover all key diagnostic lines (C IV, He II, Mg II,

Hβ, [O III]) for a large statistical sample up to z~4, and the rest frame UV diagnostic lines at the redshifts currently explored by JWST (z>4) for an even larger sample. This requires broad wavelength coverage from the optical into the NIR (ideally ~0.36-2.4 μm).

♦ **Multi-epoch spectroscopy for a tier-1 subsample** (few ×$10^2$-$10^3$ objects), with a cadence of ~7-10 days (observed frame, well matched to the expected rest-frame continuum and BLR lags and to track changes in the disc-corona state) over ≥5 years (to get a few tens of epochs per year per source), enabling ensemble reverberation mapping of continuum and broad lines. This will allow us to map continuum-line lags for Hβ, Mg II, C IV, He II, yielding ensemble BLR stratification and R-L relations as functions of $\Delta\alpha_{ox}$, $M_{BH}$, $\lambda_{Edd}$ and redshift.

♦ **Contemporaneous or quasi-simultaneous X-ray spectroscopy for the tier-1 sample** (e.g. visits within ±1-2 days of optical/NIR epochs), and sparser X-ray monitoring (few snapshots per year) for the broader population, exploiting future X-ray missions in the 2030s-2040s (e.g. Athena and successors). This will allow us to track line profile changes and outflow signatures as the coronal state evolves, providing direct evidence for how the nebular regions respond to SED variations.

♦ Beyond the BLR and NLR, a wide-field IFU with arcmin-scale fields will allow **spatially resolved mapping of CGM emission** (Lyα, CIV, HeII, [O III] etc.) for a subsample of quasars across different accretion states, linking small-scale disc-corona physics to kpc-100 kpc gas reservoirs. With NIR coverage up to the K band, the CGM can be mapped in Hα, preferable to Lyα as a non-resonant, purely recombination line that more directly traces changes in the ionising SED, out to cosmic noon. At z ~ 2.5, a 1' FOV corresponds to ~500 kpc, comfortably enclosing 100 kpc-scale nebulae.

Pathfinder works with current facilities already demonstrate the potential of this approach (e.g. Timlin+20): for example, we analysed a sample of >5000 SDSS DR16 quasars cross-matched with Chandra Source Catalog 2.0, and found emerging correlations between X-ray spectral properties and UV/optical emission-line diagnostic (with rest-frame coverage varying with redshift) as well as links between UV/optical properties and $\Delta\alpha_{ox}$ across the entire population (Fig. 2 Bisogni et al., in submission). These results hint at a continuous shift in AGN properties across the Lx-Luv plane. However, surveys that can be carried out with current and near-future instruments do not offer the necessary combination of sample size, spectral resolution, redshift coverage and cadence to fully exploit these trends. Current multi-object spectrographs, with typical multiplex values of a few $10^3$

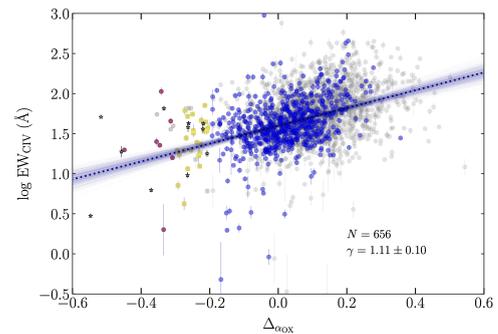

Fig. 2: Equivalent width of CIV line as a function of $\Delta\alpha_{ox}$ for a selection of normal (blue), and X-ray weak ($\Delta\alpha_{ox}$<-0.2 and <-0.3, yellow and red) CSC2.0/SDSS DR16 sources (Bisogni et al. in submission)

over <1 deg² fields, make it impractical to obtain $10^4$-$10^5$ AGN spectra with multi-epoch coverage and uniform depth within a reasonable survey duration. A transformative experiment, ensemble continuum and line reverberation mapping across parameter space, requires a dedicated survey facility with a multiplex of order $10^4$ over a wide field of view (>1 deg²), and with the capability to coordinate observations with future X-ray missions.